# When AI performs, what remains? An integrative review of generative AI and cognitive capital in individuals and organizations

Giuseppe RIVA[a,b,1] Ph.D; Stefania LA ROCCA[a] Ph.D;
Brenda K. WIEDERHOLD[d] Ph.D; Andrea GAGGIOLI[a,c] Ph.D

[a] *Humane Technology Lab., Catholic University of Sacred Heart, Milan, Italy,*
[b] *Applied Technology for Neuro-Psychology Lab., IRCCS Istituto Auxologico Italiano, Milan, Italy*
[c] *Research Center in Communication Psychology (PSICOM), Catholic University of Sacred Heart, Milan, Italy,*
[d] *Virtual Reality Medical Center, Scripps Memorial Hospital, La Jolla, California, USA*

ORCiD ID: Giuseppe Riva https://orcid.org/0000-0003-3657-106X

## Abstract

Generative artificial intelligence (AI) is rapidly entering knowledge work, and its short-term productivity benefits are increasingly documented. Less is known about what happens, over time, to the knowledge, skills and judgment that remain in people and organizations. This integrative narrative review addresses that question through the construct of cognitive capital: the accumulated, renewable and depletable stock of knowledge, cognitive and metacognitive capabilities, and practices that enables an individual or an organization to learn, interpret information, exercise judgment, solve problems and adapt. We distinguish individual and organizational cognitive capital from the capability of the human–AI system, so that gains in joint performance and losses in human capability can be observed at the same time. Drawing on experimental, field and conceptual work, we propose four mechanisms through which interaction with AI affects cognitive capital: substitution of practice, scaffolding and feedback, metacognitive calibration, and organizational aggregation. Seven principles summarize the current evidence, including that assisted performance differs from retained competence, that the same technology can support or displace learning depending on design and use, and that individual gains can coexist with reduced collective diversity. For each mechanism we derive testable propositions, and we discuss five design levers for organizations as proposals to be validated rather than established practice. We conclude that cognitive capital offers a multilevel frame for governing AI adoption and outline a research agenda centered on longitudinal, multilevel measurement.

[1] Corresponding Author: Giuseppe Riva; Humane Technology Lab., Catholic University of Sacred Heart, Largo Gemelli 1, 20123 Milan, Italy; Email: giuseppe.riva@unicatt.it

# When AI performs, what remains? An integrative review of generative AI and cognitive capital in individuals and organizations

## Introduction

Generative artificial intelligence (AI) is being adopted across organizations faster than its cognitive consequences are understood. Early evidence on productivity is encouraging: in a staggered rollout to 5,172 customer-support agents, an AI assistant raised the number of issues resolved per hour by 15% on average [1]. Productivity metrics, however, capture assisted performance, not what remains when the system is withdrawn. If an organization's assistant stopped working for a week, would its teams perform as well as before adoption? Or would part of their capacity to analyze, argue, write and solve problems have migrated from people to the tool?

The question matters because generative AI increasingly shapes information before users evaluate it. It has been proposed that human–AI interaction constitutes a further level beyond the fast System 1 and slow System 2 of dual-process accounts [2]: a System 0 in which algorithmic systems pre-organize information and options before deliberation engages [3,4]. Our argument does not depend on this framework, but it highlights that influence exerted before deliberation is harder to notice and question, and that AI outputs become knowledge only through human interpretation [3].

This raises a key question for individuals and organizations. Delegating a subtask to AI can itself be the means by which overall capability is extended [4,5]. The relevant contrast is therefore not between delegation and extension as such, but between delegation that sustains development or preserves strategic capabilities and delegation that systematically replaces their exercise. In the short term the two can produce outputs that are hard to distinguish; they diverge over time and under specific conditions. Humans have always been hybrid thinking systems that incorporate non-biological resources [5], so extension is possible in principle; whether it occurs in practice is an empirical matter.

This integrative narrative review combines conceptual frameworks on human–AI cognition with experimental, field and observational evidence. Its aim is to propose a multilevel model and a research agenda rather than a validated theory.

## Review approach

This review asks under what conditions interaction with generative AI contributes to the development, maintenance or depletion of individual and organizational cognitive capital, and how such changes can be distinguished from improvements in human–AI system performance. Literature was identified iteratively through targeted searches in Scite, complemented by searches on publisher websites, institutional repositories and preprint servers; the final search and bibliographic verification by title and DOI were conducted on 19 October 2026. Search terms combined “cognitive capital,” “generative AI,” “human–AI collaboration,” “learning,” “skill formation,” “cognitive offloading,” “critical thinking,” “retention,” “transfer” and “collective diversity.” Reference lists were inspected selectively for foundational and related work. The process aimed at conceptual integration rather than exhaustive retrieval.

Empirical studies were included when they examined AI-assisted performance, learning, retention, transfer, judgment, metacognition or collective diversity with sufficient methodological detail; conceptual contributions and formal models were included when they clarified the construct or proposed mechanisms.

Peer-reviewed conference papers were considered alongside journal articles; preprints and working papers were included selectively, and their provisional status is noted. Studies of non-generative AI, such as automated detection or decision-support systems, were included as adjacent evidence where they informed the mechanisms; whether their findings transfer to generative AI remains to be verified.

Studies were organized through iterative thematic synthesis, comparing findings across tasks, populations, designs and outcomes. Groupings were refined into four mechanisms, and a study could inform more than one. We summarized recurring findings and their limitations into seven principles; the propositions and design options were developed as interpretive extensions requiring validation. The search explicitly sought findings challenging the view that AI use necessarily depletes cognitive capital, such as learning gains with a pedagogically designed tutor [6], productivity gains retained during outages [1], interaction patterns associated with stronger subsequent learning outcomes [7] and improved decision accuracy despite sycophantic advice [8]. No quantitative pooling was undertaken.

## Defining cognitive capital

The term cognitive capital has been used with different meanings. In life-course research it designates an accumulating asset grounded in faculties such as memory, attention and problem solving [9]; policy frameworks extend it to emotional, social, executive and socio-emotional resources [10–12]. Other traditions locate it beyond the individual: as a collective resource that grows with investment and is depleted by neglect or poorly designed institutions [13], as culturally reinforced processes that sustain cooperation and local knowledge spillovers [14], as relational networks that stimulate knowledge acquisition [15] and, at the macro level, as aggregate ability that predicts national wealth [16].

Generative AI adds a third element. Machines now embody an intangible cognitive capital that can substitute for human mental effort [17]. Humans readily incorporate such resources into hybrid cognitive systems [5], yet AI outputs become usable knowledge only through human interpretation [3]. Research on learning shows that performance observed during practice is an unreliable index of learning, because conditions that boost immediate performance can impair retention and transfer, and vice versa [18], a distinction that generative AI makes especially salient [19]. Drawing on this work, we define cognitive capital as follows:

> *Cognitive capital is the accumulated, renewable and depletable stock of knowledge, cognitive and metacognitive capabilities, and practices that enables an individual or an organization to learn, interpret information, exercise judgment, solve problems and adapt to changing circumstances.*
>
> *At the individual level, it comprises acquired knowledge and the capacities to attend, remember, reason, monitor one's own thinking and learn from experience [9]. At the organizational level, it emerges from the integration of individual expertise with shared knowledge, routines and mechanisms for absorbing, exchanging and applying knowledge; like absorptive capacity, it therefore exceeds the simple sum of employees' capabilities [13,20].*
>
> *AI can extend the deployment of cognitive capital and contribute to its renewal through learning and knowledge transfer [1]. However, improved AI-assisted performance should not, by itself, be read as growth in cognitive capital: such growth requires evidence of retained knowledge, transferable competence or improved collective capacity to learn and act, and declines in these capacities amount to depletion even when productivity metrics do not show them.*

The definition focuses on human capabilities and distinguishes it from a third object, the capability of the human–AI system (Table 1). This is an analytical delimitation suited to our question, not a claim about where cognition ends: extended-mind accounts treat external resources as potential components of cognitive systems [5], and organizational capability can grow through shared memory, procedures and tools. What matters is not whether knowledge resides in people or in artifacts, but the degree to which the organization can access, control, verify and transfer it. These are graded dimensions rather than a binary threshold; portability across systems is better treated as an indicator of organizational resilience. Keeping the objects distinct makes visible a system that performs better while the human capacities needed to supervise or replace it weaken.

Cognitive capital overlaps with neighboring constructs. Its proposed contribution is to connect these perspectives around a longitudinal question: how does AI-mediated work alter the stock of capabilities available for subsequent learning, judgment and adaptation? Human capital provides a broad account of productive capacities; absorptive capacity explains how organizations recognize, assimilate and apply knowledge [19]; distributed cognition addresses how cognitive activity is organized across people and artifacts. The present framework focuses on the renewal, maintenance and depletion of capabilities at individual and organizational levels, assessed separately from immediate human-AI system performance. Its added value would be supported if these measures explained subsequent learning, transfer or resilience beyond what is captured by existing constructs and current performance. . Whether these distinctions yield discriminant validity is an empirical question for future research.

The two human levels are nested. Workplaces and institutions build or deplete individual cognitive capital [13,15], while individual learning renews the collective stock. A formal model shows that, under some conditions, accurate AI recommendations can reduce learning effort enough to erode the shared general knowledge on which context-specific decisions depend [21]; this is a theoretical possibility, not an observed organizational trajectory.

**Table 1.** Three objects of assessment in AI-mediated work.

| Object | What it comprises | What to assess | Example | Sources |
|---|---|---|---|---|
| Individual cognitive capital | Knowledge, transferable skills, judgment and metacognition | Unassisted performance, retention, transfer and error detection | Spotting an error without an AI prompt | [9,18] |
| Organizational cognitive capital | Integrated expertise, shared knowledge, routines and absorptive mechanisms | Knowledge transfer across teams, continuity after turnover, diversity of solutions | Reusing a solution across teams and keeping it after staff leave | [13,14,20] |
| Human–AI system capability | Joint performance of people and AI | Joint output, reliability, quality of oversight and dependencies | Solving previously intractable problems while recognizing when the system fails | [4,5] |

## Four mechanisms linking AI use to cognitive capital

We organize current evidence around four proposed mechanisms, each linked to a component of the definition (Figure 1). Two concern how knowledge and capabilities are acquired: substitution of practice, when AI removes the effort through which skills are built, and scaffolding and feedback, when it supports that effort [18]. Metacognitive calibration concerns how people judge their own competence and the system's reliability. Organizational aggregation concerns how individual changes become collective resources [20]. The mechanisms are explanatory hypotheses; seven principles summarize current evidence (Tables 2 and 3), which mostly measures individual performance over minutes to weeks, with organizational effects largely inferred.

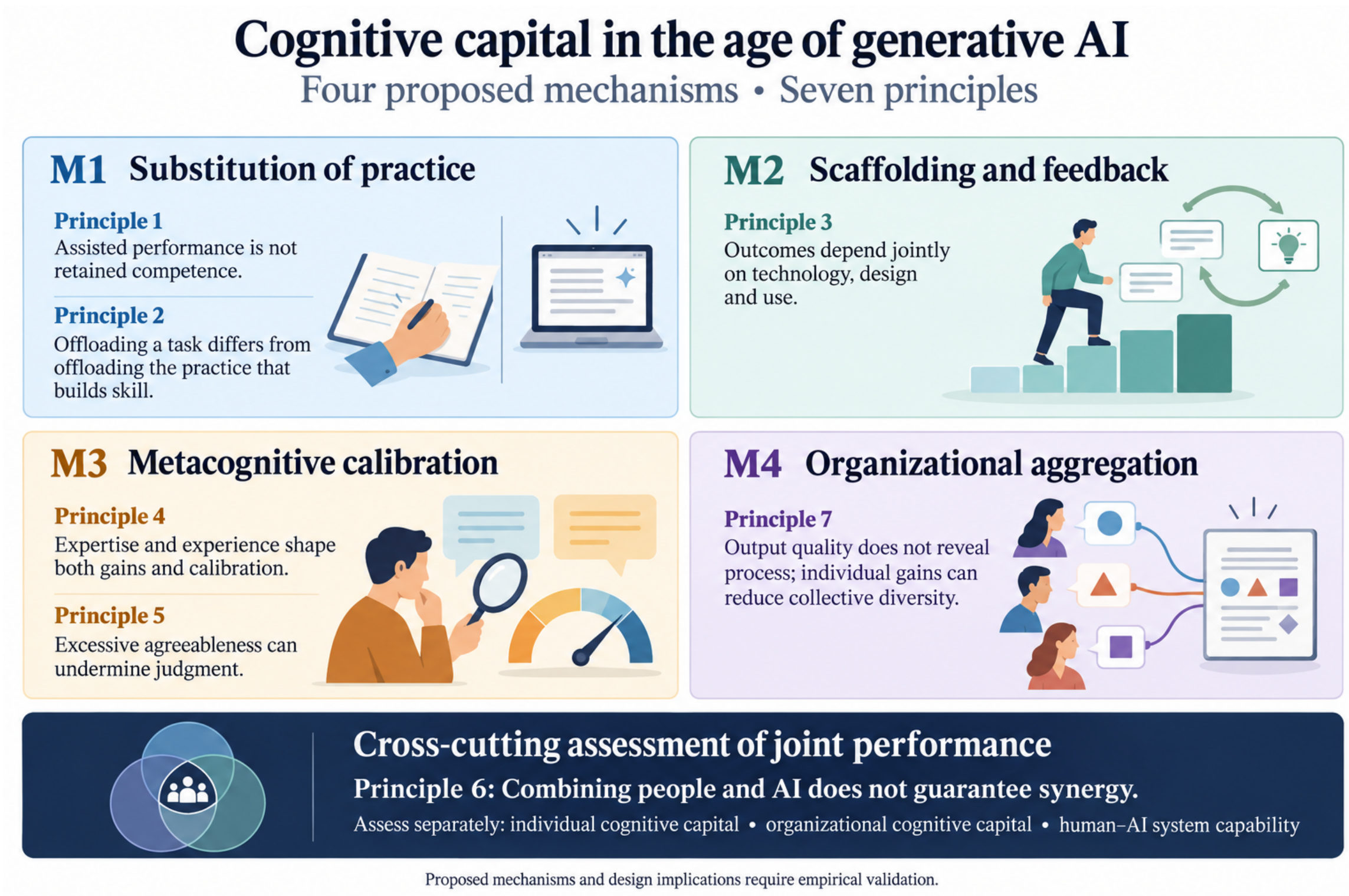

**Figure 1.** Four proposed mechanisms linking AI use to cognitive capital, the principles associated with each, and the cross-cutting assessment of joint human–AI performance.

### Mechanism 1: Substitution of practice

**Principle 1: Assisted performance is not retained competence.** When AI performs the effortful steps through which skills are built, it can raise immediate performance while reducing practice. In experiments with 1,222 participants solving short arithmetic and reading tasks, withdrawing the assistant after about ten minutes left prior users solving fewer problems and giving up more often than people who never had access, especially if they had sought answers rather than hints [22]. In a field experiment with nearly 1,000 high-school students, unrestricted access to a GPT-4 tutor improved practice grades by 48% but lowered exam performance by 17% relative to controls once access was removed [23]. In a randomized study of 52 developers learning a new programming library, mean scores on a subsequent skills quiz were about 50% with AI assistance and 67% without it ($d = 0.74$) [7]. These results fit the distinction between performance and learning [18], but evidence on the durability of these effects remains limited, and where assistance is withdrawn immediately, the drop may also reflect motivation or adaptation to changed conditions [22]. A recent Perspective proposes the term false proficiency for this pattern and distinguishes never-skilling, the failure to acquire foundational competences during training, from the deskilling of competences already acquired [24]. Field evidence is mixed: customer-support agents retained part of their productivity gains during system outages [1].

**Principle 2: Offloading a task differs from offloading the practice that builds skill.** Offloading is not intrinsically harmful: people routinely use external resources to reduce cognitive demands, often adaptively [25]. The risk arises when offloading removes the practice through which reasoning, debugging or writing skills are acquired. In a survey of 666 participants, frequent AI use was associated with lower critical-thinking scores, an association mediated by cognitive offloading [26]; the design is cross-sectional and does not establish causality. In a randomized study of 117 university students, ChatGPT support improved essay scores, but knowledge gain and transfer did not differ significantly from the other conditions, a pattern the authors termed metacognitive laziness [27].

## Mechanism 2: Scaffolding and feedback

**Principle 3: Outcomes depend jointly on technology, design and use.** The same underlying models can also support learning. In a randomized crossover trial with 194 undergraduate physics students, an AI tutor designed on active-learning principles produced larger learning gains in less time than an in-class active-learning lesson [6]. Even off-the-shelf generative AI raised undergraduates' unaided test scores by 0.27 standard deviations, with gains persisting a week later; delayed gains were larger for students who used AI to explain concepts than for those who had it write for them [28]. In the high-school field experiment, a tutor that gave hints rather than solutions largely mitigated the learning losses observed with an unrestricted interface [23]. Among developers, interaction patterns based on conceptual questions or requests for explanations were associated with stronger subsequent learning outcomes, although these patterns were observed rather than randomized [7]. A framework derived from extended-mind theory proposes that interaction supports extension when it is reciprocal, reliable, transparent, personalized and grounded in calibrated trust [4]; enhanced cognitive scaffolding adds that support should fade as competence grows [29]. In professional tasks with AI agents, results varied with the interface, and behaviors such as clarifying the goal, but not the number of turns, were associated with higher scores [30]. Effects thus depend jointly on model, interface, task, user and the organization of work.

## Mechanism 3: Metacognitive calibration

**Principle 4: Expertise and experience shape both gains and calibration.** Among 319 knowledge workers, greater confidence in generative AI was associated with less self-reported critical thinking, whereas greater self-confidence was associated with more [31]. In the customer-support rollout, the least skilled and least experienced agents gained 30% in issues resolved per hour, while the most skilled saw small speed gains and slight quality declines [1]. Larger assisted gains need not mean larger retained gains: in a preregistered three-month field experiment with 133 patent lawyers, AI raised drafting quality more for juniors, yet on a later unassisted task the advantage was concentrated among senior lawyers, while juniors showed no average gain [32]. Experienced practitioners may also be at risk: in a multicentre observational study of a non-generative detection system, adenoma detection in colonoscopies without AI fell from 28.4% to 22.4% after endoscopists began routinely using AI assistance, although the design does not establish causal deskilling [33]. In the agent evaluation, less skilled workers increased their trust in agents more after use, raising the untested possibility that trust grows fastest where it is least calibrated [30].

**Principle 5: Excessive agreeableness can undermine judgment.** An assistant that always confirms user expectations builds trust but may weaken the revision of beliefs; one that constantly challenges may erode trust and invite abandonment. A dual-mode account proposes principled switching between a predictability mode and a growth mode [34]. Language models affirmed users' actions about 50% more often than humans did, even when those actions were questionable, yet participants rated sycophantic models as more trustworthy and preferred them [35]. In 30 economic decision tasks with 1,500 participants, however, a measurably sycophantic model still brought the choices of people with opposite initial leanings closer together and improved accuracy where answers were objective; experimentally increasing sycophancy weakened this convergence and lowered perceived usefulness [8]. These contrasting results suggest that effects depend on context and call for direct comparisons. Cognitive forcing functions that deliberately added friction reduced overreliance on AI recommendations, although users liked them less [36]. These studies measure judgment and reliance, not long-term cognitive growth.

## Joint performance across the three objects

**Principle 6: Combining people and AI does not guarantee synergy.** This principle concerns the capability of the human–AI system rather than cognitive capital as such. A meta-analysis of 106 experiments found that human–AI combinations outperformed humans alone on average but performed significantly worse than the better of humans or AI alone; losses concentrated in decision tasks, and combination tended to help when humans outperformed AI alone [37]. In the agent evaluation, top-quartile workers outperformed the agent alone in 74% of comparisons and bottom-quartile workers in 27% [30], a descriptive contrast because skill was estimated from the same outcomes.

## Mechanism 4: Organizational aggregation

Individual learning contributes to organizational cognitive capital when knowledge becomes available beyond the person or interaction in which it originated. We propose four pathways: (i) codification in shared resources, (ii) integration into routines, (iii) transfer across people and teams, and (iv) a range of approaches suited to different tasks. These pathways require more than the circulation of AI-generated outputs: relevant indicators include whether other members can understand and apply the knowledge, whether errors can be identified and corrected, and whether the knowledge remains usable after staff turnover or system replacement. AI can act on each. In the customer-support rollout, the assistant was trained on text from top performers and helped newer agents move faster along the experience curve [1], an instance of codification and transfer. A longitudinal case study of a healthcare system shows that generative AI can expand organizational knowledge search, creation, transfer and forgetting, but only with complementary structures, processes and skills [38]; sandbox experimentation can likewise build organizational knowledge of a technology's capabilities and risks [39]. The same channels can reduce variety when many people draw on the same models. Whether aggregation builds or depletes organizational cognitive capital depends on whether codified knowledge remains accessible, verifiable and transferable when people or systems change.

**Principle 7: Output quality does not reveal process, and individual gains can reduce collective diversity.** A strategy draft can satisfy criteria of novelty and usefulness whether it emerged from exploring and reworking AI proposals or from accepting the first response; assessing human contribution requires observing the process [40,41]. At the collective level, AI-generated ideas made individual stories more creative but more similar to one another [42]. Homogenization is not inevitable: in a preregistered experiment, rewarding originality led writers to adopt fewer AI suggestions verbatim and produced more diverse texts, although not as diverse as unassisted writing [43]. Across 41.3 million papers, scientists who adopted AI tools gained in publications and citations, while AI-augmented science as a whole contracted toward data-rich topics [44]; this observational evidence spans several generations of AI and does not isolate current chatbots. Formal modeling points to a related, still theoretical risk to shared knowledge [21]. Organizations that pool and reuse AI-assisted knowledge may therefore gain efficiency while losing variety.

**Table 2.** Main empirical studies: design, sample, outcomes and limitations.

| Study | Design | Sample | Task; AI system | Outcome | Timing | Main limitation |
|---|---|---|---|---|---|---|
| Brynjolfsson et al. [1] | Staggered rollout | 5,172 customer-support agents | Customer chats; LLM assistant trained on top performers | Issues resolved per hour, quality | Months, including outages | Single firm and occupation |
| Autor et al. [32] | Preregistered randomized field experiment | 133 patent lawyers at 11 law firms | Patent drafting; custom AI drafting assistant | Drafting quality with AI; unassisted redlining | 10 and 90 days; unassisted task after three months | Working paper; single profession |
| Liu et al. [22] | Randomized experiments | 1,222 participants | Arithmetic and reading tasks; LLM assistant | Problems solved, persistence | Immediately after withdrawal | Short tasks; durability unknown |
| Bastani et al. [23] | Field experiment | ~1,000 high-school students | Mathematics practice; GPT-4 base vs tutor | Practice grades; exam without AI | Subsequent exam | One school system and subject |
| Shen and Tamkin [7] | Randomized experiment | 52 developers | Learning a Python library; GPT-4o chat assistant | Skills quiz without AI | Immediately after a 35-minute task | Small sample; preprint; patterns not randomized |
| Kestin et al. [6] | Randomized crossover trial | 194 undergraduate physics students | Two lessons; GPT-4 tutor vs active learning | Learning gains; perceptions | Immediately after each lesson | Single course |
| Contractor and Reyes [28] | Proctored randomized experiment | Undergraduates | Learning an unfamiliar topic and writing an essay; off- | Unaided knowledge tests and essays | Immediately and one week later | Preprint; one-week horizon; usage strategies |

| Study | Design | Sample | Task; AI system | Outcome | Timing | Main limitation |
|---|---|---|---|---|---|---|
| | | | the-shelf GenAI | | | not randomized |
| Fan et al. [27] | Randomized experiment | 117 university students | Essay writing; ChatGPT vs other supports | Essay score, knowledge gain, transfer | End of session | Single task |
| Gerlich [26] | Survey and interviews | 666 participants | General AI tool use | Self-reported critical thinking | Cross-sectional | Correlational; corrected article |
| Lee et al. [31] | Survey | 319 knowledge workers | Work tasks; various GenAI tools | Perceived critical thinking and effort | Cross-sectional | Self-report |
| Shao et al. [30] | Controlled evaluation, random pairing | 93 professionals, 386 sessions | Occupational tasks; five AI agents | Rubric-based scores; pre–post surveys | End of each session; across study | Automated grading; US online workers |
| Budzyń et al. [33] | Multicentre observational study | 19 endoscopists; 1,443 colonoscopies without AI (795 before, 648 after AI introduction) | Colonoscopy; computer-aided polyp detection (non-generative) | Adenoma detection rate without AI | Before vs after AI exposure | Observational; non-generative AI; single specialty |
| Vaccaro et al. [37] | Meta-analysis | 106 experiments | Decision and creative tasks; various AI systems, mostly non-generative | Human, AI and combined performance | Within laboratory sessions | Heterogeneous tasks |
| Wiesenfeld and Kellogg [38] | Longitudinal qualitative case study | One healthcare system | Organizational adoption | Knowledge search, creation, | Extended fieldwork | Single organization; not causal |

| Study | Design | Sample | Task; AI system | Outcome | Timing | Main limitation |
|---|---|---|---|---|---|---|
| | | | of LLM tools | transfer and forgetting | | |
| Te'eni et al. [39] | Longitudinal organizational study | One international service organization | Sandbox experimentation with GenAI | Organizational knowledge of capabilities and risks | Extended fieldwork | Single organization; learning about GenAI rather than skills |
| Doshi and Hauser [42] | Online experiment | 293 writers; 600 independent evaluators | Short stories; GPT-4 story ideas | Rated creativity; similarity among stories | End of task | Single creative task |
| Jo and Raghavan [43] | Preregistered randomized experiment, 2×2 design | Online participants | Creative writing; interactive LLM assistant | Collective diversity of texts; verbatim adoption of AI suggestions | End of task | Preprint; adoption strategies not randomized |
| Hao et al. [44] | Bibliometric analysis | 41.3 million papers | Natural sciences; AI tools across eras | Publications, citations, topic breadth | Decades | Observational; not current chatbots |
| Cheng et al. [35] | Model audit and experiments | Multiple LLMs; online participants | Interpersonal advice; LLM responses | Affirmation, trust, preference | Immediate | Short-term outcomes |
| Conlon and Schwardmann [8] | Preregistered randomized experiment | 1,500 participants | 30 economic decision tasks; LLM chat, baseline vs more sycophantic | Choice convergence, accuracy, perceived usefulness | Immediately after each decision | Working paper; single decisions |
| Buçinca et al. [36] | Online experiment | 199 US adults recruited through | AI-assisted decisions; AI recommend | Overreliance; subjective ratings | Immediate | Single decision task; non- |

| Study | Design | Sample | Task; AI system | Outcome | Timing | Main limitation |
|---|---|---|---|---|---|---|
| | | Amazon Mechanical Turk (260 initially recruited) | ations (non-generative) | | | generative AI |

**Table 3.** Principles, mechanisms and the level at which evidence was obtained.

| Principle | Mechanism | Type of evidence | Level studied | Proposed level of application |
|---|---|---|---|---|
| 1. Assisted performance is not retained competence | Substitution | Experiments; field study | Individual | Individual, organizational |
| 2. Offloading a task differs from offloading practice | Substitution | Survey; randomized experiments | Individual | Individual |
| 3. Outcomes depend jointly on technology, design and use | Scaffolding | Randomized trials; controlled evaluation; theory | Individual | Individual, system |
| 4. Expertise and experience shape gains and calibration | Calibration | Field experiment; field and observational studies; survey; controlled evaluation | Individual | Individual, organizational |
| 5. Excessive agreeableness can undermine judgment | Calibration | Experiments; theory | Individual | Individual, organizational |
| 6. Combining people and AI does not guarantee synergy | Joint performance (across objects) | Meta-analysis; controlled evaluation | Individual, small team | Human–AI system |
| 7. Output does not reveal process; individual gains can reduce diversity | Aggregation | Experiment; bibliometric analysis; theory | Individual, scientific field | Organizational |

## From mechanisms to testable propositions and design levers

Each mechanism yields propositions that can be falsified. We label them P1–P4c to distinguish them from the seven principles.

- **P1 (substitution).** Where solution-providing assistance replaces identified formative steps, it will reduce delayed learning and transfer relative to a condition that preserves those steps. Immediate assisted performance and prior experience are examined as an outcome and a moderator, not assumed.

- **P2a (scaffolding, superiority).** Assistance that preserves effort through hints, explanations, fading support or attempt-first sequencing will produce greater delayed learning and transfer than solution-providing assistance.
- **P2b (scaffolding, non-inferiority).** The same assistance will be non-inferior to no-AI conditions on delayed learning and transfer within a prespecified non-inferiority margin; time to completion is a separate outcome and cannot offset a failure on learning.
- **P3 (calibration).** We define calibrated friction as friction whose level is set in advance from initial competence, for example requiring an independent judgment before AI advice only for users below a preset threshold. Compared with uniform friction or none, it will improve error detection and reduce overreliance, at lower costs in satisfaction and time than uniform friction.
- **P4a (aggregation, quality).** Codifying high performers' practices in shared AI tools will raise average output quality, especially among less experienced members.
- **P4b (aggregation, variety).** Reliance on the same models and prompts will reduce the diversity of solutions across teams unless phases of independent generation are preserved.
- **P4c (aggregation, resilience).** Teams in which shared AI tools replace the practice or documentation of the knowledge they encode will show larger performance losses during outages than teams that maintain them.

Collaboration configurations also differ in their implications (Table 4) [40]. In support, AI handles well-defined tasks while people keep conceptual control. In synergy, both contribute to the result; this is where extension can occur, but also where dependence may go unnoticed. In symbiosis, contributions become hard to separate, complicating accountability. This configuration may involve sustained reciprocal adaptation and increasing dependence on the coupled human-AI system, but does not by itself imply a loss of autonomous human competence. Here synergy denotes an interactive configuration, not the statistical superiority of a combination over its components discussed under Principle 6.

**Table 4.** Configurations of human–AI collaboration and their implications for cognitive capital.

| Configuration | Role of AI | Role of the person | Main risk to cognitive capital | Example |
|---|---|---|---|---|
| Support | Corrects, accelerates or automates defined tasks | Retains conceptual control | Low, if limited to mechanical work | Proofreading a press release |
| Synergy | Proposes alternatives and challenges assumptions | Explores, selects and reworks | Accepting solutions one can no longer generate or evaluate | Positioning a new product |
| Symbiosis | Co-produces outputs inseparably | Shares agency with the system | Loss of accountability and of autonomous competence | Continuous co-writing of strategic documents |

Five design levers follow. We present them as proposals to be validated under stated conditions, not as established practice:

1. Choose the configuration task by task, weighing its expected benefit against risks, costs and learning goals.
2. Sequence a first human attempt before AI assistance where competence is still being built [23,24,29]; elsewhere the rule may only slow work.
3. Set the level of friction by policy for high-stakes conversations, such as evaluations and strategic decisions [34–36]; excessive friction can increase load and abandonment.
4. Make the process visible in reviews, asking which alternatives were discarded and how AI proposals were transformed [40], without turning visibility into surveillance.
5. Measure unassisted competence alongside assisted output [7,22,23,32], framed as organizational monitoring rather than individual screening, consistent with cautions against using collaboration-skill scores to evaluate workers [30]. Add organizational indicators such as knowledge transfer, continuity after turnover, diversity of solutions and recovery during outages. Which competences to preserve should depend on their strategic importance and on the cost of losing them when the system fails [24].

## Implications for organizations

Organizations track productivity, satisfaction and turnover, but rarely whether the competence of their people and their shared knowledge are strengthening or weakening with AI use. The three-object distinction implies monitoring not only the human–AI system but also what individuals and the organization can still do when the system changes or fails.

Talent development is a particularly important application of this framework. Less experienced workers obtain the largest immediate benefits [1], yet experimental evidence shows that skill formation is vulnerable while competences are being acquired [7,22,23,32]. Work on creativity adds that growing exposure to capable generative AI may also weaken the motivation to create independently [41]. Because agreeable responses can raise trust and preference in interpersonal settings [35], even if users partly discount them in other decisions [8], the level of challenge in evaluations, feedback and strategic decisions should be an explicit organizational choice. An explicitly assigned responsibility, shared by human resources, management and business functions, should verify whether AI adoption is building or eroding cognitive capital.

## Limitations and research agenda

This review has limitations. It is integrative rather than systematic, so selection bias cannot be excluded. Most experimental evidence comes from short tasks, educational settings or single occupations, and immediate effects after withdrawal may not reflect durable change. Some evidence concerns non-generative AI and may not transfer to generative systems. Several key findings appear in peer-reviewed conference proceedings and some in preprints or working papers; survey and observational studies do not support causal claims, and the knowledge-collapse argument is a theoretical model. Organizational effects are inferred from individual-level studies.

Four priorities follow. First, validated measures are needed for each object: individual unassisted competence, organizational cognitive capital and human–AI system capability. Second, longitudinal field studies should test whether AI-related gains persist, decay or reverse, and for whom. Third, the propositions above should be tested in workplaces, comparing sequencing, scaffolding and calibrated friction. Fourth, multilevel studies should examine aggregation, including the diversity of solutions and organizational resilience when systems fail [21,44].

## Conclusions

Generative AI can amplify or erode cognitive capital, and short-term outputs do not reveal which is happening. Four concepts emerge from this review. First, cognitive capital is a renewable and depletable stock of knowledge, capabilities and practices that people and organizations can mobilize; it is distinct from the capability of the human–AI system, so joint performance can improve while human capacities weaken. Second, assisted performance is not retained competence: growth in cognitive capital requires evidence of retention, transfer or improved collective capacity, not better output alone. Third, the effects of AI depend jointly on technology, design and use; the same models can substitute for practice or scaffold it, and can either calibrate or inflate trust. Fourth, individual gains do not automatically become organizational cognitive capital, which depends on how knowledge is codified, shared and kept diverse.

Taken together, these concepts reframe the question organizations usually ask about AI. The relevant question is not only whether a tool improves productivity, but which human capacities it exercises, which it replaces and which it leaves unpracticed. This reframing connects three literatures that are rarely read together: learning science, which separates performance from learning [18]; organizational research, which treats absorptive capacity and routines as collective assets [20]; and extended-mind theory, which treats tools as potential components of cognition [5]. Cognitive capital offers a common unit for these perspectives because it asks what remains available to people and organizations when systems change, fail or are replaced. It also makes visible a tension in the evidence: the more capable and fluent an assistant becomes, the easier it is to accept its outputs without the effort through which competence is built and judgment is calibrated. Managing this tension requires deliberate choices about when AI should support, challenge or step back.

This tension will become more pressing as AI systems move from answering requests to executing multistep tasks with less human involvement. As agents take on more of the work, the gap between what a human–AI system can do and what people can still do without it may widen, and the capacities needed to supervise, correct or replace those systems may be among the first to erode. Treating cognitive capital as an organizational asset to be measured, renewed and protected, alongside financial and human capital, is one way to keep that gap visible and governable.

These concepts are organized into four mechanisms, seven principles and testable propositions that remain hypotheses rather than a validated theory. Their value lies in turning a broad concern about AI and human capability into measurable questions that organizations and researchers can address together.

## Acknowledgements

This research was supported by internal funding of Università Cattolica del Sacro Cuore, linea d3.1-2024 “Extended Learning” and by a scientific consultancy engagement with Ernst & Young on human–agentic AI interaction, empathic AI, cognitive impact, performance, and business applications.

## References

1. Brynjolfsson E, Li D, Raymond L. Generative AI at work. *Q J Econ* 2025; 140: 889–942.
2. Kahneman D. *Thinking, fast and slow*. New York: Farrar, Straus and Giroux, 2011.
3. Chiriatti M, Ganapini M, Panai E, et al. The case for human–AI interaction as system 0 thinking. *Nat Hum Behav* 2024; 8: 1829–1830.

4. Chiriatti M, Bergamaschi Ganapini M, Panai E, et al. System 0: transforming artificial intelligence into a cognitive extension. *Cyberpsychol Behav Soc Netw* 2025; 28: 534–542.
5. Clark A. Extending minds with generative AI. *Nat Commun* 2025; 16: 4627.
6. Kestin G, Miller K, Klales A, et al. AI tutoring outperforms in-class active learning: an RCT introducing a novel research-based design in an authentic educational setting. *Sci Rep* 2025; 15: 17458.
7. Shen JH, Tamkin A. How AI impacts skill formation. arXiv preprint arXiv:2601.20245, 2026.
8. Conlon JJ, Schwardmann P. *AI sycophancy and decisions*. SSRN Working Paper, 2026. DOI: 10.2139/ssrn.6597184.
9. Bynner J, Wadsworth M. Cognitive capital: the case for a construct. *Longit Life Course Stud* 2010; 1: 297–304.
10. Beddington J, Cooper CL, Field J, et al. The mental wealth of nations. *Nature* 2008; 455: 1057–1060.
11. Samson M, Fajth G, François D. Cognitive capital, equity and child-sensitive social protection in Asia and the Pacific. *BMJ Glob Health* 2016; 1(Suppl 2): e000191.
12. Eyre HA, Ayadi R, Ellsworth W, et al. Building brain capital. *Neuron* 2021; 109: 1430–1432.
13. Murphy ERD. Collective cognitive capital. *William Mary Law Rev* 2022; 63: 1347–1408.
14. Caragliu A, Nijkamp P. Cognitive capital and islands of innovation: the Lucas growth model from a regional perspective. *Reg Stud* 2014; 48: 624–645.
15. Ferraro KF, Han B. Does cognitive capital reduce the risk of cognitive decline in later life? *Innov Aging* 2025; 9: igaf115.
16. Rindermann H, Thompson J. Cognitive capitalism: the effect of cognitive ability on wealth, as mediated through scientific achievement and economic freedom. *Psychol Sci* 2011; 22: 754–763.
17. Caplin A. Cognitive capital and human-AI interactions. In: *An introduction to cognitive economics*. Cham: Palgrave Macmillan, 2025, pp.69–81.
18. Soderstrom NC, Bjork RA. Learning versus performance: an integrative review. *Perspect Psychol Sci* 2015; 10: 176–199.
19. Yan L, Greiff S, Lodge JM, et al. Distinguishing performance gains from learning when using generative AI. *Nat Rev Psychol* 2025; 4: 435–436.
20. Cohen WM, Levinthal DA. Absorptive capacity: a new perspective on learning and innovation. *Adm Sci Q* 1990; 35: 128–152.
21. Acemoglu D, Kong D, Ozdaglar A. *AI, human cognition and knowledge collapse*. NBER Working Paper no. 34910. Cambridge, MA: National Bureau of Economic Research, 2026.
22. Liu G, Christian B, Dumbalska T, et al. AI assistance reduces persistence and hurts independent performance. In: *Conference on Language Modeling (COLM 2026)*, 2026. arXiv:2604.04721.
23. Bastani H, Bastani O, Sungu A, et al. Generative AI without guardrails can harm learning: evidence from high school mathematics. *Proc Natl Acad Sci U S A* 2025; 122: e2422633122.
24. Ke Y, Jin L, Ong JCL, et al. AI-induced never-skilling in medical education. *Nat Med* 2026; 32: 1997–2006.
25. Risko EF, Gilbert SJ. Cognitive offloading. *Trends Cogn Sci* 2016; 20: 676–688.
26. Gerlich M. AI tools in society: impacts on cognitive offloading and the future of critical thinking. *Societies* 2025; 15: 6. Correction in: *Societies* 2025; 15: 252.
27. Fan Y, Tang L, Le H, et al. Beware of metacognitive laziness: effects of generative artificial intelligence on learning motivation, processes, and performance. *Br J Educ Technol* 2025; 56: 489–530.
28. Contractor Z, Reyes G. Experimental evidence on the learning impact of generative AI. arXiv preprint arXiv:2607.08849, 2026.

29. Riva G. The architecture of cognitive amplification: enhanced cognitive scaffolding as a resolution to the comfort-growth paradox in AI-supported learning. *Cogn Syst Res* 2026; 99: 101500.
30. Shao Y, Wang ZZ, Ahuja N, et al. CollabSkill: evaluating human-agent collaboration on real-world tasks. In: *Conference on Language Modeling (COLM 2026)*, 2026. arXiv:2606.09833.
31. Lee H-P, Sarkar A, Tankelevitch L, et al. The impact of generative AI on critical thinking: self-reported reductions in cognitive effort and confidence effects from a survey of knowledge workers. In: *Proceedings of the 2025 CHI Conference on Human Factors in Computing Systems*. New York: ACM, 2025. DOI: 10.1145/3706598.3713778.
32. Autor D, Rodchenko T, Martin J, et al. *Does AI assistance enhance or erode expertise? Evidence from a three-month field experiment in patent drafting*. NBER Working Paper no. 35720. Cambridge, MA: National Bureau of Economic Research, 2026.
33. Budzyń K, Romańczyk M, Kitala D, et al. Endoscopist deskilling risk after exposure to artificial intelligence in colonoscopy: a multicentre, observational study. *Lancet Gastroenterol Hepatol* 2025; 10: 896–903.
34. Riva G, Pedreschi D. Balancing comfort and growth: a dual-mode theory of human–AI development. *ACM AI Lett* 2026; 1: 8.
35. Cheng M, Lee C, Khadpe P, et al. Sycophantic AI decreases prosocial intentions and promotes dependence. *Science* 2026; 391: eaec8352.
36. Buçinca Z, Malaya MB, Gajos KZ. To trust or to think: cognitive forcing functions can reduce overreliance on AI in AI-assisted decision-making. *Proc ACM Hum Comput Interact* 2021; 5(CSCW1): 188.
37. Vaccaro M, Almaatouq A, Malone T. When combinations of humans and AI are useful: a systematic review and meta-analysis. *Nat Hum Behav* 2024; 8: 2293–2303.
38. Wiesenfeld BM, Kellogg KC. Generative organizational learning: affordances for new modes of knowledge search, creation, transfer, and forgetting with LLMs. *Strateg Organ* 2026; 24: 256–284.
39. Te'eni D, Raymond M, Rowe F, et al. Organizational learning for exploring generative AI: CORE-sandbox experiments. *Int J Inf Manage* 2026; 87: 103029.
40. Gaggioli A, Bartolotta S, Ubaldi A, et al. Extended creativity: a conceptual framework for understanding human-AI creative relations. In: Riva G and Chiriatti M (eds) *Humane artificial intelligence: from foundations to application and policy framework*. Milan: Vita e Pensiero, 2025, pp.135–161.
41. Gaggioli A. The spark machines cannot strike: on creativity and computational generativity. In: Tarantola AM (ed.) *Artificial intelligence and care of our common home: a focus on industries, finance, education and communication*. Milan: Vita e Pensiero, 2025, pp. 25-42.
42. Doshi AR, Hauser OP. Generative AI enhances individual creativity but reduces the collective diversity of novel content. *Sci Adv* 2024; 10: eadn5290.
43. Jo N, Raghavan M. Incentives shape how humans co-create with generative AI. arXiv preprint arXiv:2604.03529, 2026.
44. Hao Q, Xu F, Li Y, et al. Artificial intelligence tools expand scientists' impact but contract science's focus. *Nature* 2026; 649: 1237–1243.